\documentclass[3p,times,twocolumn]{elsarticle}

\usepackage{ecrc}
\usepackage{hyperref} 
\usepackage{float}
\usepackage{color}
\newcommand{\cen}[1]{\multicolumn{1}{c}{#1}}

\newcommand{\av}[1]{\langle #1 \rangle}

\volume{00}

\firstpage{1}

\journalname{Nuclear and Particle Physics Proceedings}

\runauth{F. Mahmoudi et al.}

\jid{nppp}

\jnltitlelogo{Nuclear and Particle Physics Proceedings}

\usepackage{amssymb}

\biboptions{sort&compress}

\usepackage[figuresright]{rotating}

\begin{document}

\begin{frontmatter}



\dochead{}

\title{Present Status of $b \to s \ell^+ \ell^-$ Anomalies\tnoteref{label1}}
\tnotetext[label1]{Based on the talk by F.M. at the Sixth Workshop on Theory, 
Phenomenology and Experiments in Flavour Physics, Capri, June 2016.}


\author{Farvah Mahmoudi}

\address{Univ Lyon, Univ Lyon 1, ENS de Lyon, CNRS, Centre de Recherche Astrophysique de Lyon\\
UMR5574, F-69230 Saint-Genis-Laval, France;\\
Theoretical Physics Department, CERN, CH-1211 Geneva 23, Switzerland}
\ead{nazila@cern.ch}

\author{Tobias Hurth}

\address{PRISMA Cluster of Excellence and Institute for Physics (THEP)\\
Johannes Gutenberg University, D-55099 Mainz, Germany}
\ead{tobias.hurth@cern.ch}

\author{Siavash Neshatpour}

\address{School
 of Particles and Accelerators, Institute for Research in Fundamental Sciences (IPM)\\
P.O. Box 19395-5531, Tehran, Iran}
\ead{neshatpour@ipm.ir}

\begin{abstract}
We discuss the observed deviations in $b \to s \ell^+ \ell^-$ processes from the Standard Model predictions and present global fits for the New Physics description of these anomalies.
We further investigate the stability of the global fits under different theoretical assumptions and suggest strategies and a number of observables to clear up the source of the anomalies.
\end{abstract}


\end{frontmatter}


\section{Introduction}
\label{sec:intro}
The angular observables of the $B \to K^* \mu^+ \mu^-$ decay were measured in 2013 by the LHCb collaboration using 1 fb$^{-1}$ of data~\cite{Aaij:2013qta}.
While most of the measured observables agreed with their Standard Model (SM) predictions, the angular observable $P^\prime_5$ was in 3.7$\sigma$ tension with respect to the SM prediction, in the $q^2 \in [4.30, 8.63]$ GeV$^2$ bin.
Assuming the anomaly is not due to an underestimation of the hadronic effects or a statistical fluctuation in
the experimental data, global analysis of  $b\to s$ data  indicated that a New Physics (NP) explanation for this anomaly is in the effective theory language a reduction of about 25\% in the Wilson coefficient $C_9$~\cite{Descotes-Genon:2013wba,Altmannshofer:2013foa,Beaujean:2013soa,Hurth:2013ssa,Mahmoudi:2014mja}.
In 2014, LHCb presented experimental results for the ratio $R_K\equiv BR(B^+\to K^+ \mu^+ \mu^-)/BR(B^+\to K^+ e^+ e^-)$ in the $q^2 \in [1,6]$ GeV$^2$ bin which was in 2.6$\sigma$ tension with the SM prediction~\cite{Aaij:2014ora}. This anomaly which points toward violation of lepton flavour universality can be explained through a reduction in $C_9^\mu$ which is consistent with the NP explanation for the $P^\prime_5$ anomaly~\cite{Hurth:2014vma,Altmannshofer:2014rta}. This observable, unlike the $B \to K^* \mu^+ \mu^-$ observables is theoretically very clean.
Furthermore, in 2015 the LHCb collaboration measured a number of $B_s \to \phi \mu^+ \mu^-$ 
observables where the branching ratio in the $q^2 \in [1,6]$ GeV$^2$ bin is in tension with the SM prediction at 3.2$\sigma$~\cite{Aaij:2015esa}, which again can be explained with a reduction in $C_9$~\cite{Hurth:2014vma}.
In 2015, LHCb updated the measurements of $B \to K^* \mu^+ \mu^-$ observables with 3 fb$^{-1}$ of data where the tension in $P_5^\prime$ (in the $[4,6]$ and $[6,8]$ GeV$^2$ bins) remained although with slightly less significance~\cite{Aaij:2015oid}. In addition, recently the Belle experiment presented a measurement of the angular observables of $B \to K^* \ell^+ \ell^-$ ~\cite{Abdesselam:2016llu} which shows a 2$\sigma$ deviation for $P^\prime_5$ in the $q^2 \in [4,8]$ GeV$^2$ bin, which supports the LHCb result.

For the exclusive decays $B_{(s)} \to K^*(\phi) \mu^+ \mu^-$, where the final state meson is a vector meson, 
the long-distance contributions of the electromagnetic and semileptonic operators ($O_{7,9,10}$) 
can be described through  seven independent form factors $V, A_{0,1,2},T_{1,2,3}$.
The form factors are usually considered a main source of uncertainty as non-perturbative calculations are required to estimate them.
There are further hadronic effects from the four-quark and chromomagnetic operators ($O_{1-6}$ and $O_8$, respectively) accompanied with the exchange of a virtual photon. The matrix elements of these operators cannot all be factorised into form factors, giving rise to non-factorisable corrections.
At low $q^2$, in the heavy quark and large energy limit, these effects are calculable at leading order in $\Lambda/m_b$  in QCD factorisation and its field-theoretical formulation of Soft-Collinear Effective Theory (SCET).
However, higher powers of the non-factorisable effects are not known and until further calculations
become available can only be ``\emph{guesstimated}'' (a partial calculation of the power corrections for the $B\to K^* \ell^+ \ell^-$ decay is available through a phenomenological description~\cite{Khodjamirian:2010vf}). 

In the low $q^2$ region  the  seven a priori independent form factors can be reduced to two 
soft form factors $\xi_{\perp,\parallel}$~\cite{Charles:1998dr}, up to corrections of ${\cal O}$($\alpha_s$) and 
${\cal O}$($1/m_b$), and while the former corrections have been calculated the latter remain unknown.
These unknown factorisable power corrections can  be guesstimated through dimensional arguments or by fitting ad hoc functions when comparing with the full form factors~\cite{Descotes-Genon:2014uoa}.
Reduction of the seven form factors to two soft form factors makes it possible to construct angular observables which are (soft) form factor independent at leading order~\cite{Egede:2008uy,Egede:2010zc}.
One set of such form factor independent observables includes the so-called optimised ($P_i^\prime$) observables.
These observables are specially interesting in the absence of correlations among the form factor uncertainties, 
since otherwise there can be an overestimation of the theoretical errors.

The two theoretical strategies for $B_{(s)} \to K^*(\phi) \ell^+ \ell^-$ decays are therefore using the soft form factors (soft FF) or the full form factors (full FF).
In either of the approaches the significance of the anomalies is dependent on the estimated size of the power corrections.
For the soft FF approach, we estimate the factorisable and non-factorisable power corrections collectively by 
varying the tranversity amplitudes according to
\[A_{\perp,\parallel,0} \to A_{\perp,\parallel,0} \times \left( 1+ b_i e^{\theta_i} + c_i (q^2/6 \; {\rm  GeV}^2)e^{\phi_i} \right),\]
where $i$ stands for $\perp,\parallel,0$, and with $\theta_i,\phi_i\in (-\pi,\pi)$ and
$b_i \in (-0.1,+0.1)$, $(-0.2,+0.2)$, $(-0.3,+0.3)$ ranges and  
$c_i \in (-0.25,+0.25)$, $(-0.5,+0.5)$, $(-0.75,+0.75)$ ranges which in the following we  
refer to as 10, 20 and 30\% error for the power corrections, respectively. 
In the full FF approach only the non-factorisable power corrections are relevant
which we consider by multiplying the hadronic part of the tranversity amplitudes  
by a multiplicative factor similar to the soft FF case with
$b_i \in (-0.05,+0.05), (-0.1,+0.1), (-0.2,+0.2),(-0.6,+0.6)$ and 
$c_i \in (-0.125,+0.125)$, $(-0.25,+0.25)$, $(-0.5,+0.5)$, $(-1.5,+1.5)$ for the 5, 10, 20 and 60\% error, respectively.

\section{Model-independent New Physics fits}

We perform global fits using {\tt SuperIso v3.6}~\cite{Mahmoudi:2007vz,Mahmoudi:2008tp,Mahmoudi:2009zz} by calculating and minimising the $\chi^2$ in which all the theoretical and experimental correlations are considered~\cite{Hurth:2016fbr}. 

Assuming New Physics to appear only in one operator, a model independent analysis of all the relevant $b\to s $ data 
favours NP models with negative contributions to $C_9$. The best fit values for various one operator
fits are given in Tab.~\ref{tab:OneOpFit}, where the full FF approach with 10\% power correction error has been considered for the theoretical prediction of the $B{(s)}\to K^*(\phi) \mu^+ \mu^-$ observables.
\begin{table}[h!]
\begin{center}
\scalebox{0.80}{
\begin{tabular}{l|cccc}
 & \cen{b.f. value} &  ${\rm Pull}_{\rm SM}$ & 68\% C.L. &95\% C.L. \\ 
\hline \hline
%
& & & & \\[-3.mm] 
$\delta C_{9}/C_{9}^{\rm SM} $            & $-0.18$  & $3.0\sigma$  & $[-0.25,-0.09]$  & $[-0.30,-0.03]$   \\[1.mm] 
$\delta C_{9}^\prime/C_{9}^{{\rm SM}} $   & $+0.03$  & $1.0\sigma$  & $[-0.05,+0.12]$  & $[-0.11,+0.18]$   \\[1.mm]
$\delta C_{10}/C_{10}^{\rm SM} $          & $-0.12$  & $1.9\sigma$  & $[-0.23,-0.02]$  & $[-0.31,+0.04]$   \\[1.mm] \hline
& & & & \\[-3.mm] 
$\delta C_{9}^e/C_{9}^{{\rm SM}} $        & $+0.25$  & $2.9\sigma$  & $[+0.11,+0.36]$  & $[+0.03,+0.46]$   \\[1.mm]
$\delta C_{9}^{\mu}/C_{9}^{\rm SM} $      & $-0.21$  & $4.2\sigma$  & $[-0.27,-0.13]$  & $[-0.32,-0.08]$    
\end{tabular}}
\caption{Best fit values and the corresponding 68 and 95\% confidence level intervals 
in the one operator global fit to the $b \to s$ data. 
In the last two rows the fits are done when considering lepton non-universality.
\label{tab:OneOpFit}} 
\end{center} 
\end{table}

As can be seen from the table, the most probable scenario is for a reduction in $C_9^\mu$ in which the SM value is in 4.2$\sigma$ tension with the best fit value of $C_9^\mu$.

\begin{figure*}[t!]
\begin{center}
\includegraphics[width=5.5cm]{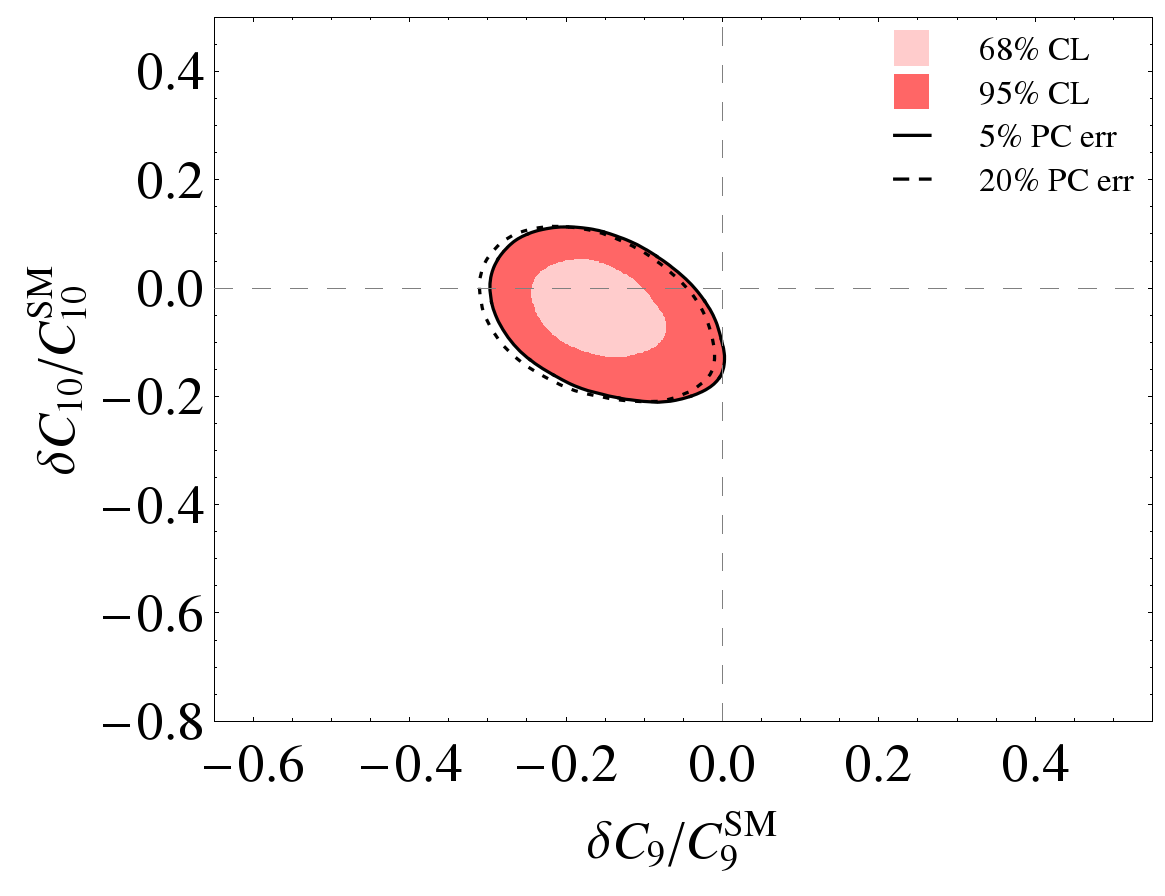}\includegraphics[width=5.5cm]{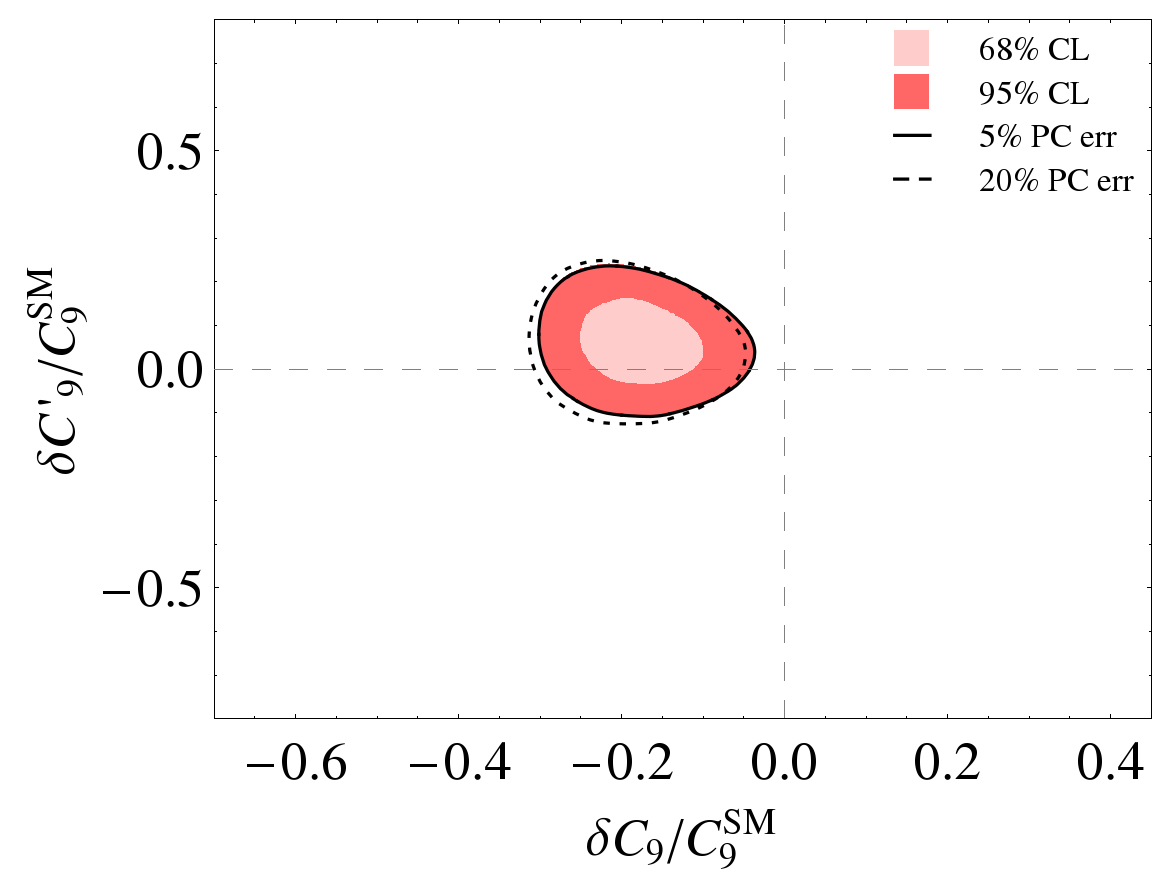}\includegraphics[width=5.5cm]{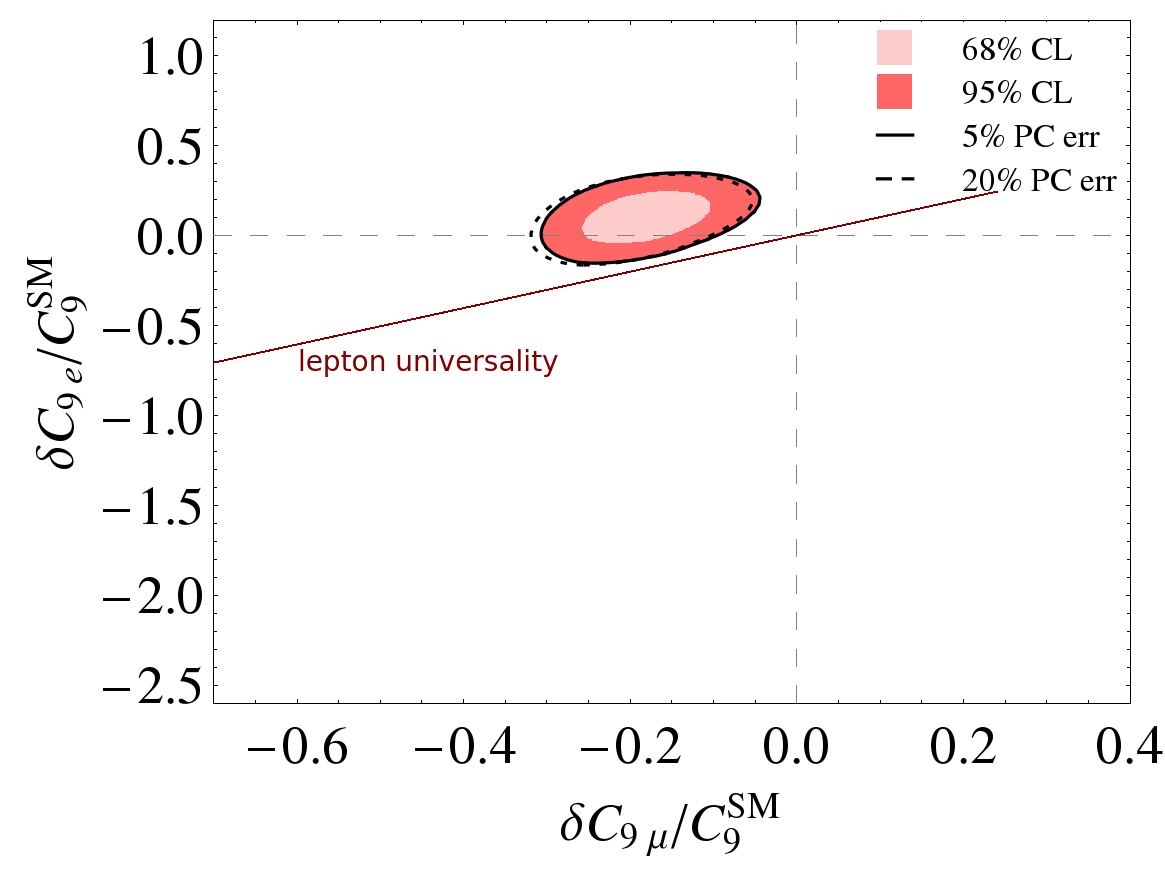}
\caption{Global fit results using the full FF approach and assuming 10\% power correction.
The solid (dashed) lines correspond to the allowed regions at 2$\sigma$ when considering 5\% (20\%) power corrections.
\label{fig:2op}}
\end{center}
\end{figure*}

It is also plausible for NP effects to appear in more than one operator.
Assuming two operator fits, the results of the fits for the \{$C_{9},C_{10}$\}, \{$C_{9},C_{9}^\prime$\} and \{$C_{9}^e,C_{9}^\mu$\} sets have been shown in Fig.~\ref{fig:2op} where we considered the full FF approach with 5, 10 and 20\% power correction errors. In all cases $\delta C_9$ has more than 2$\sigma$ tension with the SM. 
From the left plot in Fig.~\ref{fig:2op} it is obvious that the best description of the data is when assuming lepton flavour violation, what can also be seen in the last line of Tab.~\ref{tab:OneOpFit}.

\begin{figure}[!t]
\begin{center}
\includegraphics[width=6.cm]{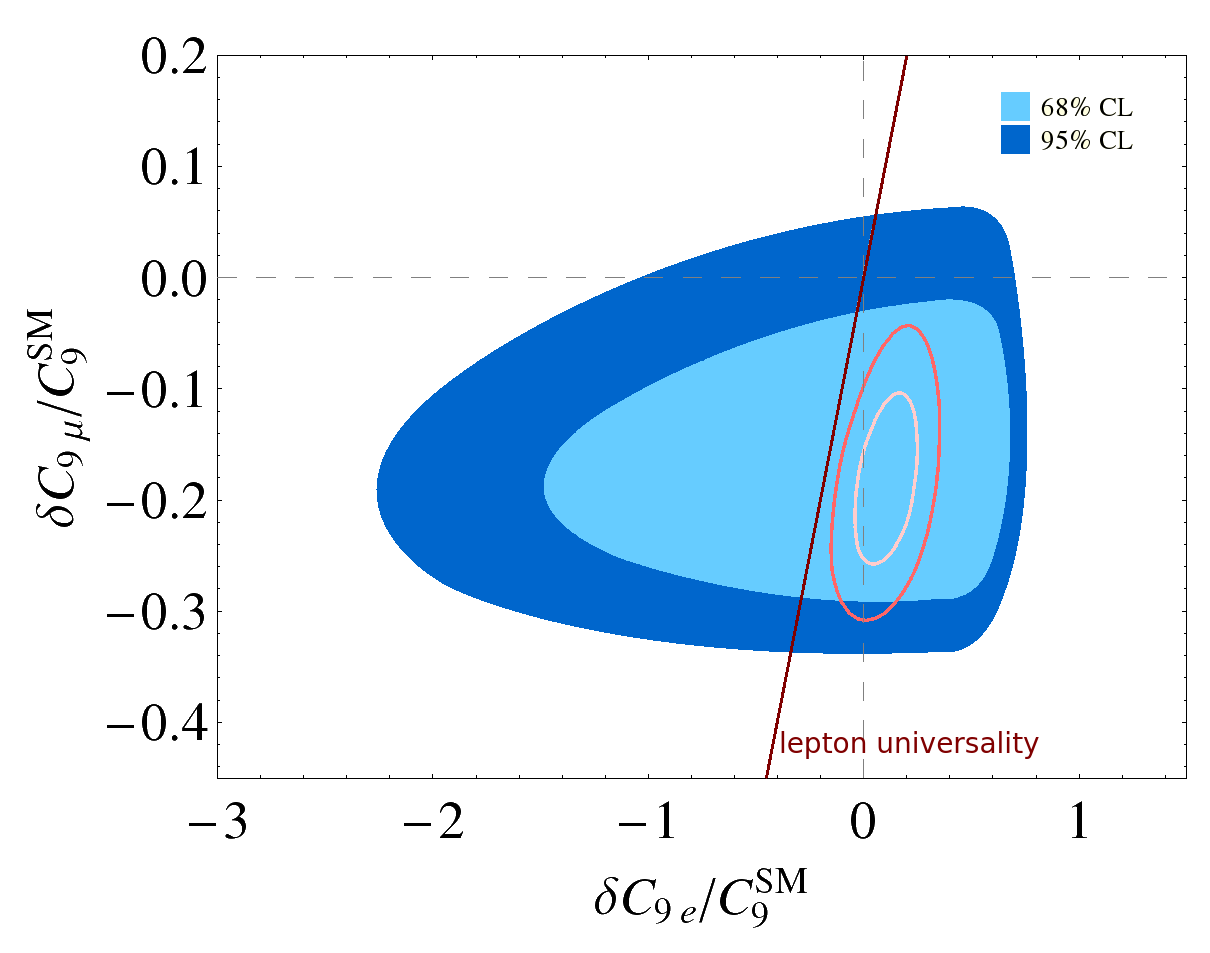}
\caption{Global fit results for \{$C_9^e, C_9^\mu, C_{10}^{e}, C_{10}^{\mu}$\} using the full FF approach and assuming 10\% power correction. 
The (light) red contour shows the (1) 2$\sigma$ allowed region when new physics is considered in two operators only, for comparison. 
The brown line corresponds to the lepton flavour universality condition.
\label{fig:4op}}
\end{center}
\end{figure}

One can go one step further and consider new physics contributions to several Wilson coefficients at the same time.
Fig.~\ref{fig:4op} shows an example of such a fit for the \{$C_9^e, C_9^\mu, C_{10}^{e}, C_{10}^{\mu}$\} set.
Allowing for NP effects in four operators, the constraints on the Wilson coefficients $C_9^e$ and $C_9^\mu$ 
can be considerably relaxed compared to the two-operator fit and the tension of $\delta C_9^{(\mu)}$ with SM can be reduced to 1$\sigma$. However, in this case the overall tension on lepton-universality is appearing in the other Wilson coefficients $C_{10}^{e}$ and $C_{10}^{\mu}$ which are not shown in the projected plane (see~\cite{Hurth:2016fbr} for more details).

\section{Stability of the fits with respect to theoretical assumptions}
\begin{figure}[!t]
\begin{center}
\includegraphics[width=6.cm]{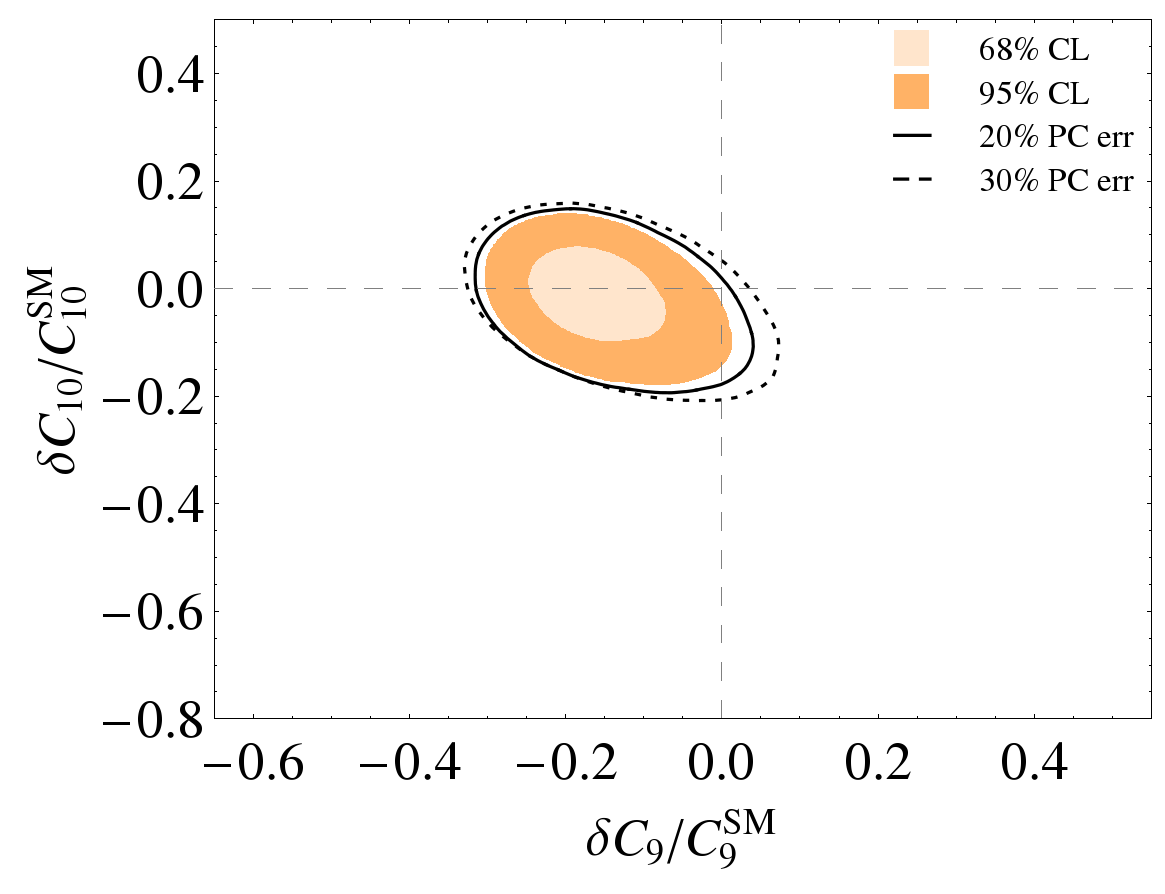}
\caption{Global fit results for \{$C_9,C_{10}$\} where the soft FF approach has been used, assuming 10\% power correction error. The solid (dashed) lines correspond to the allowed regions at 2$\sigma$ when considering 20\% (30\%)
power correction error.}
\label{fig:full_vs_soft}
\end{center}
\end{figure}

In order to check the dependence of the model-independent fit on the theoretical assumptions 
we first compare the fit results for the soft and full FF approaches. 
In Fig.~\ref{fig:full_vs_soft} the two operator fit for \{$O_9,O_{10}$\} has been shown when using the soft FF approach and assuming 10\% power correction error. Fig.~\ref{fig:full_vs_soft} should be compared with the left plot in Fig.~\ref{fig:2op}.
In both cases, regardless of the theoretical approach used, there is more than 2$\sigma$ tension for $\delta C_9$ with respect to the SM.
Moreover, in the full FF approach assuming up to 20\% power correction error, the picture remains almost the same indicating that power corrections are the sub-dominant theoretical error.

\begin{figure}[t!]
\begin{center}
\includegraphics[width=6.cm]{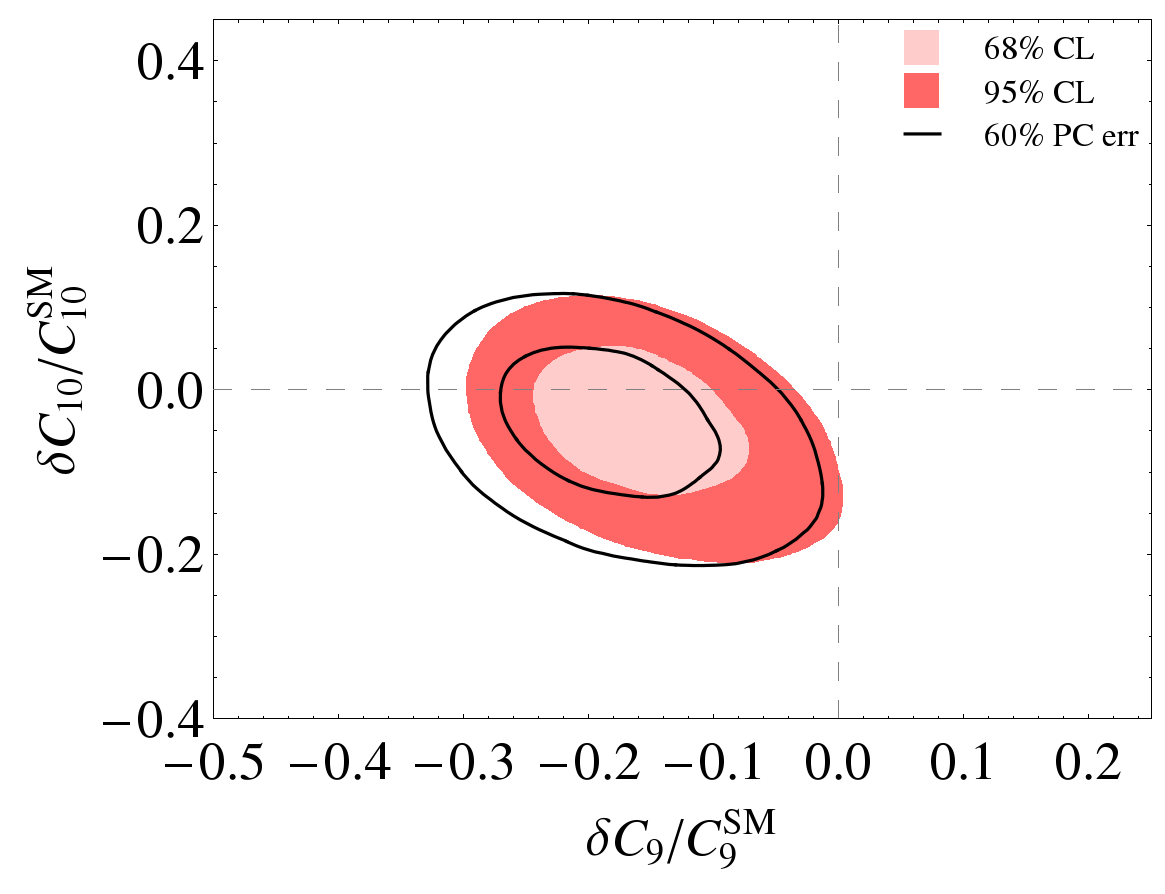}
\includegraphics[width=6.cm]{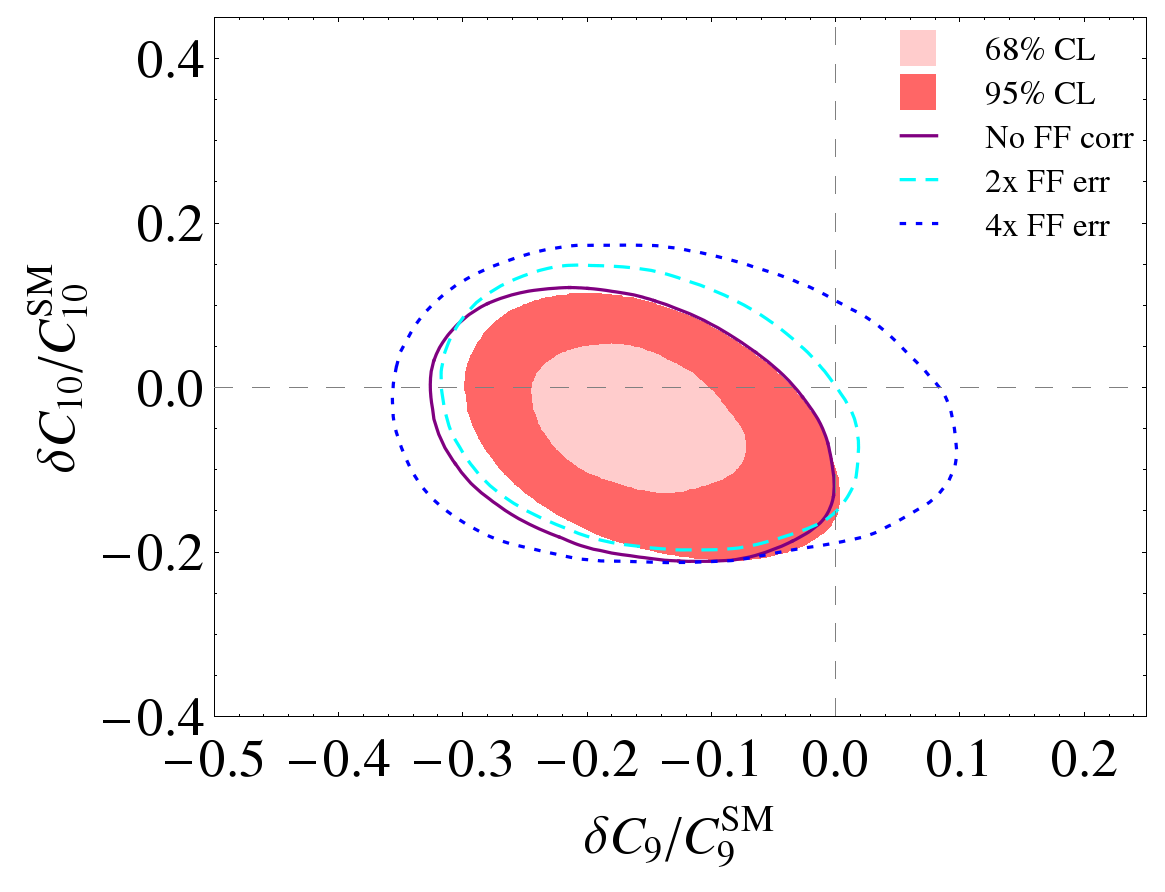}
\caption{Global fit results for \{$C_9,C_{10}$\} where the full FF approach has been used, assuming 10\% power corrections. 
On the upper plot, the solid lines correspond to the 1 and 2$\sigma$ contours when considering 60\% power correction error.
On the lower plot, the 2$\sigma$ contours when removing the form factor correlations, 
as well as when doubling (quadrupling) the form factor errors are shown with solid and dashed (dotted) lines, respectively.
\label{fig:60pc}}
\end{center}
\end{figure}

Interestingly, assuming up to 60\% error due to power corrections in 
the full FF approach the tension is not reduced (upper plot in Fig.~\ref{fig:60pc}). 
In fact a 60\% power correction error at the amplitude level results in a $\sim$20\% effect at the observable 
level.
Hence, in the full FF approach where the factorisable power corrections are included and only  non-factorisable 
power corrections are unknown, the fit results are quite stable for reasonable estimations of these corrections.
In a recent paper~\cite{Ciuchini:2015qxb}, 
instead of fitting the data to NP, a fit was done for the hadronic power corrections using a 
$q^2$ polynomial function with 18 free parameters. 
As expected the data can be explained with the hadronic fit, however, the fit results lead to corrections of up to 50\% in some critical bins at the level of observables. 
This corresponds to a power correction error of $\sim$150\% of the leading order {\it non-factorisable} contributions calculated in QCDf at the amplitude level. Such a large correction somewhat questions the validity of QCDf.\footnote{The corrections on the amplitude level in the fit to the data and the power correction error in the approach used above are two different quantities. The order of magnitude of both quantities should be similar though.}

In the lower plot in Fig.~\ref{fig:60pc}, it can be seen that the fit results are also rather  stable with respect to the form factor uncertainties when doubling them. The 2.6$\sigma$ tension of the best fit value is reduced to 2.1$\sigma$, and it is only quadrupling the form factor uncertainties which brings the tension below 2$\sigma$. For the discussion of the correlations used in the LCSR calculation of the form factors~\cite{Straub:2015ica} we refer the reader to Ref.~\cite{Hurth:2016fbr}. 

Similar results are obtained for \{$C_{9},C_{9}^\prime$\} and \{$C_{9}^e,C_{9}^\mu$\} fits 
when assuming the different mentioned theoretical assumptions (see~\cite{Hurth:2016fbr}).

\section{Stability of the fits with respect to experimental data and specific observables}

\begin{figure}[t!]
\begin{center}
\includegraphics[width=6.cm]{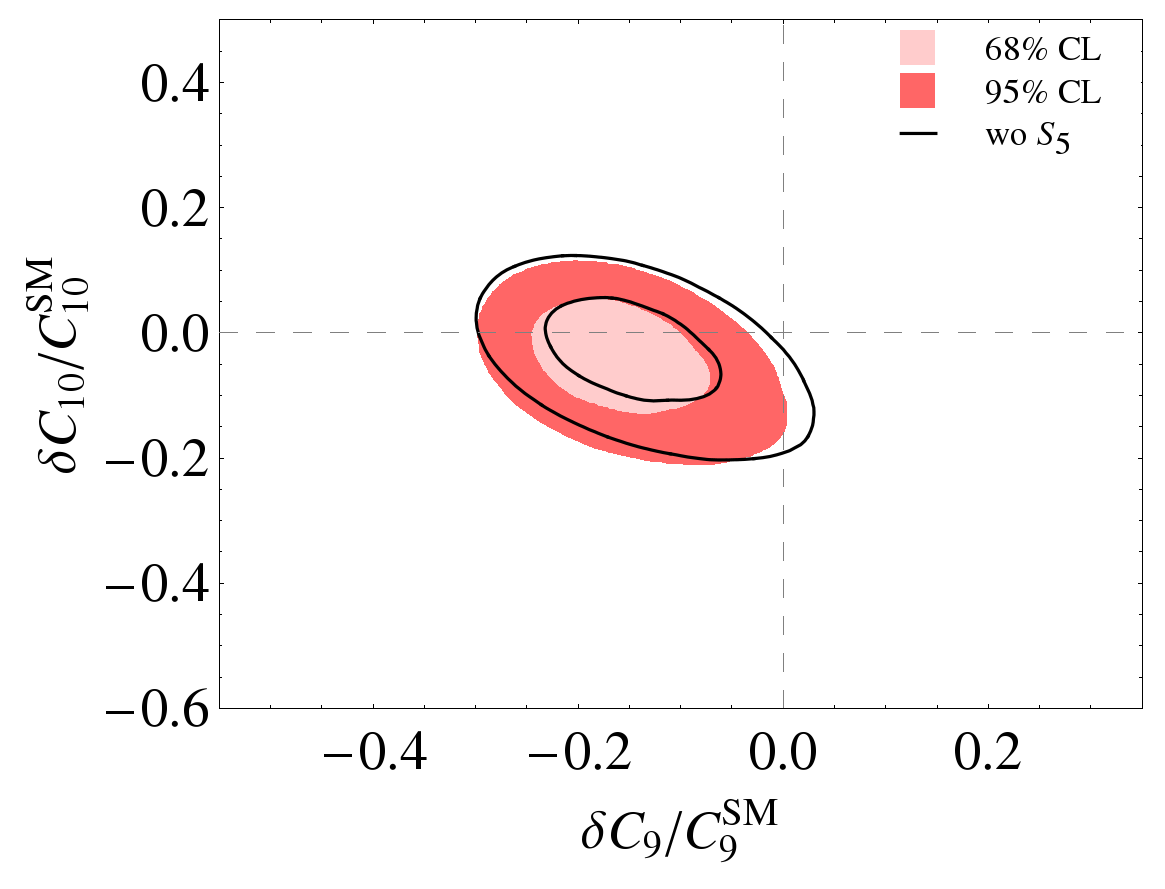}
\includegraphics[width=6.cm]{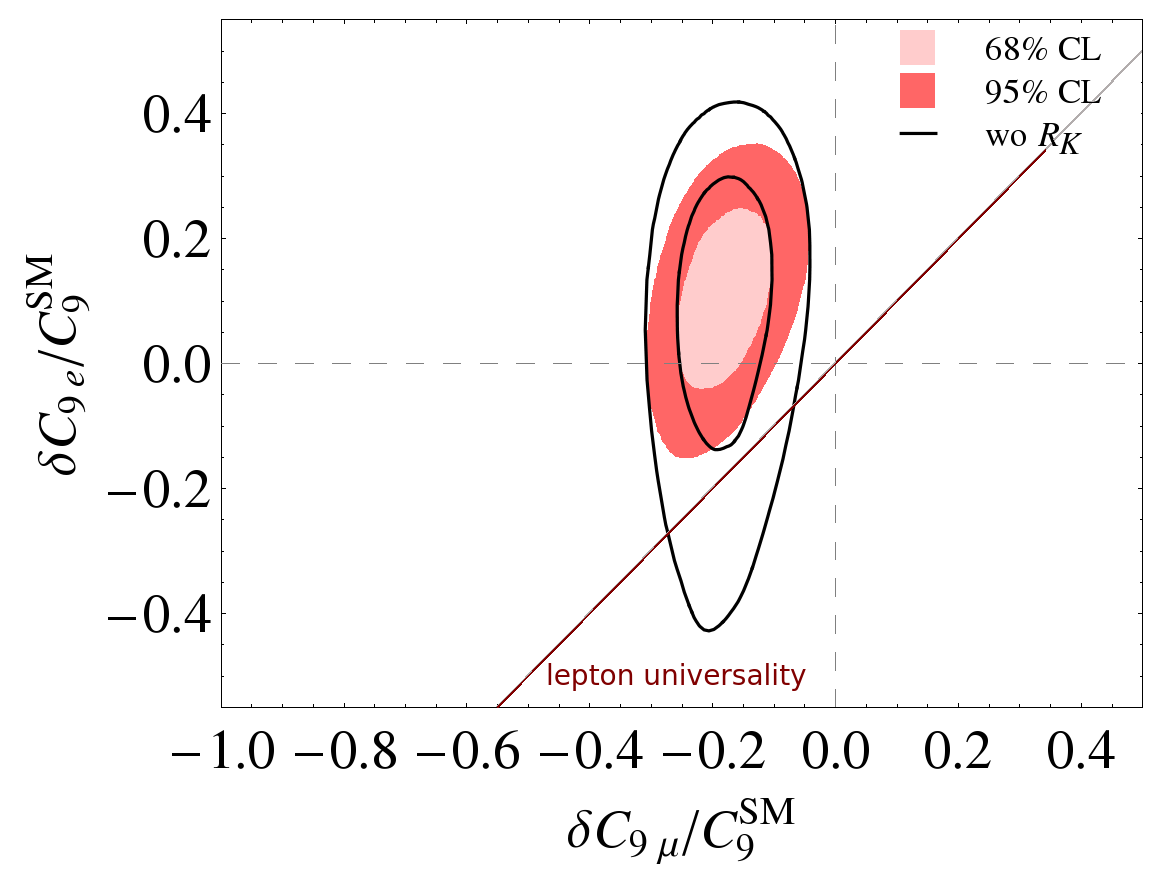}
\caption{Global fit results for \{$C_9,C_{10}$\} and \{$C_9^e,C_9^\mu$\}, using the full FF approach with 10\% power correction error. 
On the upper (lower) plot the solid lines corresponds to the 1 and 2$\sigma$ allowed regions when removing the data on the $S_5$ $(R_K)$ observable.
\label{fig:S5_RK_removed}}
\end{center}
\end{figure}

One may check also the influence of a given observable on the fit. To do so, we can remove the observable from the fit and study the consequences.
On the upper plot in Fig.~\ref{fig:S5_RK_removed}, the 1 and 2$\sigma$ regions of the \{$C_9,C_{10}$\} fit have been shown when removing the data on the $S_5$ observable (this is equivalent to removing $P_5^\prime \left(=S_5/\sqrt{F_L(1-F_L)}\right)$ for which the largest anomaly is observed).
Interestingly, even when removing the data on $S_5$, the best fit value of $\delta C_9$ is still in tension with the SM, implying that it is not only the anomaly in $S_5$ $(P_5^\prime)$ which drives $\delta C_9$ to negative values but this is rather an overall behaviour of the $b\to s$ data. 
\begin{figure}[h!]
\begin{center}
\includegraphics[width=6.cm]{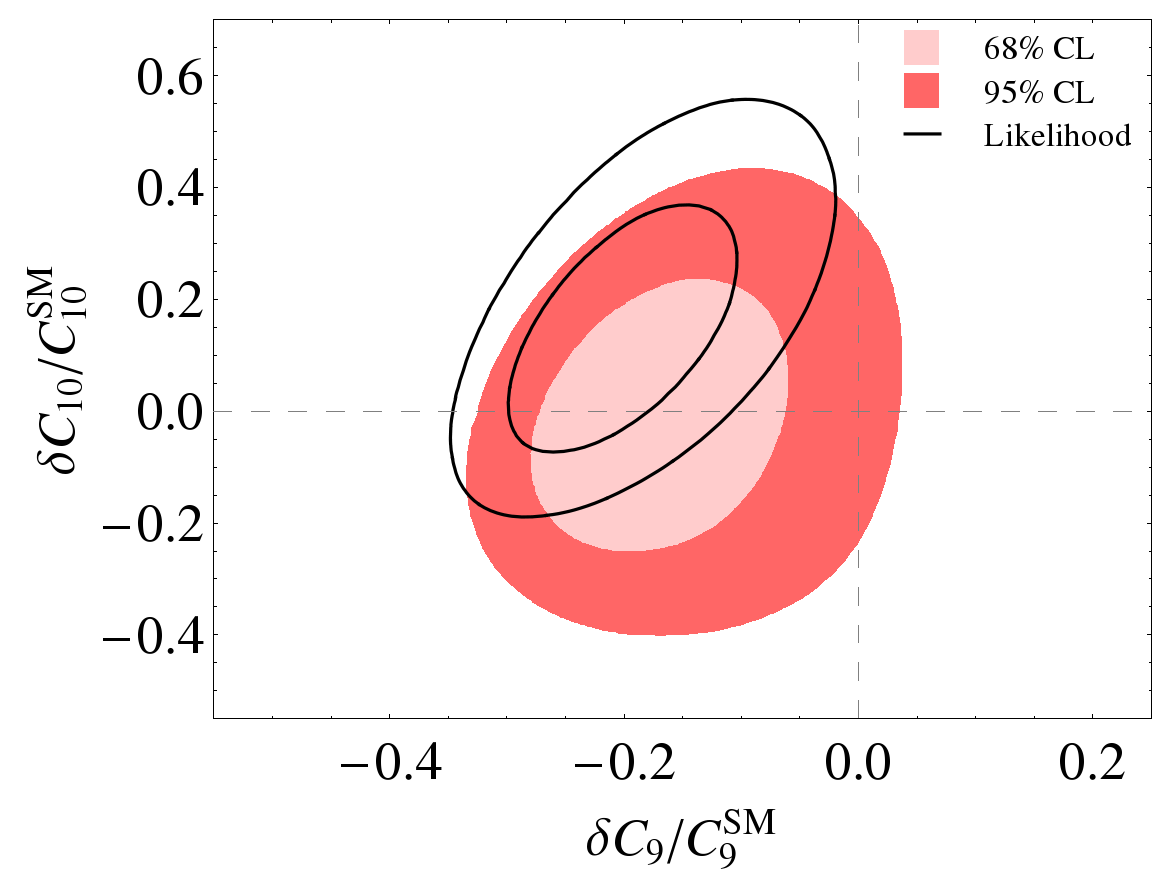}
\caption{Global fit results for \{$C_9,C_{10}$\} using only the $B\to K^* \mu^+ \mu^-$ data in the full
FF approach with 10\% power correction. The coloured regions (solid lines) correspond to the fit results using the method of moment (most likelihood) at 1 and 2$\sigma$.
\label{fig:two_exp}}
\end{center}
\end{figure}

On the other hand, the observable which drives the fit to favour lepton flavour universality violation (LFUV) is $R_K$,
since the only NP interpretation of the $R_K$ anomaly is through LFUV scenarios (where the Wilson coefficients for the muon and electron sectors are independent of each other).
This can be seen in the lower plot in Fig.~\ref{fig:S5_RK_removed} where the 1 and 2$\sigma$ allowed regions are shown when removing the data on $R_K$. In this case the tension of the best fit point with lepton flavour universality is only at $\sim1\sigma$ level.

The LHCb collaboration analysed the 3 fb$^{-1}$ data on $B\to K^* \mu^+ \mu^-$ with two different methods, the method of moments and the most likelihood method, where the former is less precise but more robust with respect to the latter.
In order to see the effect of the different methods on the fit results, the two operator fits using only the data on $B\to K^* \mu^+ \mu^-$ observables have been presented in Fig~\ref{fig:two_exp}.
The tension of the best fit point with SM is less in the case where the experimental results analysed with the method of moments have been used which is mostly due to the larger experimental errors.  
Here, again, similar results are obtained for \{$C_{9},C_{9}^\prime$\} and \{$C_{9}^e,C_{9}^\mu$\} fits (see~\cite{Hurth:2016fbr}).

\section{Strategies to identify the source of the anomalies}

In order to identify the source of the observed anomalies, several paths can be undertaken. The most important issue is due to the unknown power corrections, as the significance of the anomalies depends on the assumptions on the power corrections. Going towards an estimate of these corrections would therefore be the first strategy to disentangle the New Physics effects from the hadronic corrections. The challenge here is that these corrections are not calculable in QCDf, although alternative approaches exist based on light cone sum rule techniques~\cite{Khodjamirian:2010vf}. 
The present anomalies strongly call for such an estimate based on these techniques.

\begin{table}[!t]
\begin{center}
\scalebox{0.80}{\begin{tabular}{c|c}
Observable & Prediction \\
\hline \hline  
& \\[-3.mm] 
${\rm BR}( B\to X_s \mu^+ \mu^- )/{\rm BR}( B\to X_s e^+ e^- )_{q^2 \in[1,6] ({\rm GeV})^2}$                   &$ [0.61, 0.93] $\\ [2mm]
${\rm BR}( B\to X_s \mu^+ \mu^- )/{\rm BR}( B\to X_s e^+ e^- )_{q^2 > 14.2 ({\rm GeV})^2}$                     &$ [0.68, 1.13] $\\ [2mm]
${\rm BR}( B^0\to K^{*0} \mu^+ \mu^- )/\;{\rm BR}( B^0\to K^{*0} e^+ e^- )_{q^2 \in[1,6] ({\rm GeV})^2}$       &$ [0.65, 0.96] $\\ [2mm]
$\av{F_L(B^0\to K^{*0} \mu^+ \mu^-)}/\av{F_L(B^0\to K^{*0} e^+ e^-)}_{q^2 \in[1,6] ({\rm GeV})^2}$             &$ [0.85, 0.96] $\\ [2mm]
$\hspace*{-0.2cm}\av{A_{F\!B}(B^0\to K^{*0} \mu^+ \mu^-)}/\av{A_{F\!B}(B^0\to K^{*0} e^+ e^-)}_{q^2 \in[4,6]  ({\rm GeV})^2}$\hspace*{-0.2cm} & \hspace*{-0.2cm}$[-0.21, 0.71]$\hspace*{-0.2cm}\\ [2mm]
$\av{S_5(B^0\to K^{*0} \mu^+ \mu^-)}/\av{S_5(B^0\to K^{*0} e^+ e^-)}_{q^2 \in[4,6] ({\rm GeV})^2}$             &$ [0.53, 0.92] $\\ [2mm]
${\rm BR}( B^0\to K^{*0} \mu^+ \mu^- )/\;{\rm BR}( B^0\to K^{*0} e^+ e^- )_{q^2 \in[15,19] ({\rm GeV})^2}$ & $ [0.58, 0.95] $\\ [2mm]
$\av{F_L(B^0\to K^{*0} \mu^+ \mu^-)}/\av{F_L(B^0\to K^{*0} e^+ e^-)}_{q^2 \in[15,19] ({\rm GeV})^2}$           & \hspace*{-0.2cm}$[0.998, 0.999]$\hspace*{-0.2cm}\\ [2mm]
\hspace*{-0.2cm}$\av{A_{F\!B}(B^0\to K^{*0} \mu^+ \mu^-)}/\av{A_{F\!B}(B^0\to K^{*0} e^+ e^-)}_{q^2 \in[15,19] ({\rm GeV})^2}$\hspace*{-0.2cm} & $ [0.87, 1.01] $\\ [2mm]
$\av{S_5(B^0\to K^{*0} \mu^+ \mu^-)}/\av{S_5(B^0\to K^{*0} e^+ e^-)}_{q^2 \in[15,19] ({\rm GeV})^2}$           &$ [0.87, 1.01] $\\ [2mm]
${\rm BR}( B^+\to K^{+} \mu^+ \mu^- )/\;{\rm BR}( B^+\to K^{+} e^+ e^- )_{q^2 \in[1,6] ({\rm GeV})^2}$         &$ [0.58, 0.95] $\\ [2mm]
${\rm BR}( B^+\to K^{+} \mu^+ \mu^- )/\;{\rm BR}( B^+\to K^{+} e^+ e^- )_{q^2 \in[15,22] ({\rm GeV})^2}$       &$ [0.58, 0.95] $\\ 
\end{tabular} }
\caption{Predictions at 95\% C.L. for the ratios of $B$ decays with muons in the final state, to electrons in the final state.
\label{tab:ratios}}
\end{center} 
\end{table} 

The second strategy would be the cross check with other ratios of decays to muons versus decays to electrons, similar to the $R_K$ ratio. The advantage of such ratios is that they are free from hadronic uncertainties and the deviations cannot be explained by the unknown power corrections. In Tab.~\ref{tab:ratios} we give predictions for such ratios based on the global fit considering the two operator fit within the \{$C_9^e,C_9^\mu$\} set~\cite{Hurth:2016fbr}. It can be seen from the table that in most cases the SM point lies outside the 2$\sigma$ region of the indirect predictions. In particular, the ratio in the case of the $A_{FB}$ shows a large deviation, making it a precious observable to cross check with.\footnote{Another interesting proposal in this direction was made very recently~\cite{Serra:2016ivr}.} 

The third strategy would be to cross check with the measurements in the inclusive decay of $B\to X_s \ell^+\ell^-$, which is theoretically well-explored
. 
Predictions based on the model independent analysis show that at the future Belle II it will be possible to confirm the NP interpretation of the observed anomalies in the exclusive decays~\cite{Hurth:2013ssa,Hurth:2014zja,Hurth:2014vma}.

\section{Conclusions}

The latest LHCb results, based on the 3 fb$^{-1}$ of data set still show some tensions with the SM predictions. Model independent fits point to about 25\% reduction in the Wilson coefficient $C_9$, with New Physics in the muonic contributions being preferred. We have studied the stability of the fits both with respect to theoretical assumptions and with respect to experimental analyses, and showed that doubling the form factor errors and/or increasing the non-factorisable power corrections by up to 60\% do not change the fit results dramatically. 
Nevertheless, the LHCb data can be fitted by a general ansatz for hadronic power corrections, but  somehow huge  power corrections are needed in the fit within the critical bins.

In addition, we have identified strategies in order to understand the origin of the observed anomalies, based on the LHCb data and on the future Belle II measurements.

\section*{Acknowledgements}
We would like to thank the organisers for their invitation to the workshop, and would like to express a special thank 
to the Mainz Institute for Theoretical Physics (MITP), 
the Universit\`a di Napoli Federico II and INFN for their hospitality and support.





\nocite{*}
\bibliographystyle{elsarticle-num}
\bibliography{CapriRefs}







\end{document}